\documentclass[twocolumn,preprintnumbers,amsmath,amssymb]{revtex4}

\begin{document}


\preprint{hep-th/0609214, IFUP-TH/2006-21, TIT/HEP-559, UT-Komaba/06-9}

\title{Universal Reconnection of Non-Abelian Cosmic Strings}

%

\author{
Minoru {\sc Eto}$^{1,a}$,
Koji {\sc Hashimoto}$^{1,b}$,
Giacomo {\sc Marmorini}$^{2,3,c}$,\\
Muneto {\sc Nitta}$^{4,d}$,
Keisuke {\sc Ohashi}$^{5,e}$
and
Walter {\sc Vinci}$^{2,6,f}$
}

\affiliation{
$^1$ {\it Institute of Physics, University of Tokyo, Komaba 3-8-1,
Tokyo 153-8902, Japan}\\
$^2$ {\it INFN, Sezione di Pisa, Largo Pontecorvo, 3, Ed. C, 56127 Pisa,
Italy}\\
$^3$ {\it Scuola Normale Superiore, Piazza dei Cavalieri, 7, 56126
Pisa, Italy}\\
$^4$ {\it Department of Physics, Keio University, Hiyoshi, Yokohama,
Kanagawa 223-8521, Japan}\\
$^5$ {\it Department of Physics, Tokyo Institute of
Technology, Tokyo 152-8551, Japan}\\
$^6$ {\it Department of Physics, University of Pisa, Largo Pontecorvo, 3,
Ed. C,  56127 Pisa, Italy}\\
{\small
$^a$ {\tt meto(at)hep1.c.u-tokyo.ac.jp}, \hspace{3pt}
$^b$ {\tt koji(at)hep1.c.u-tokyo.ac.jp}, \hspace{3pt}
$^c$ {\tt g.marmorini(at)sns.it},\\
$^d$ {\tt nitta(at)phys-h.keio.ac.jp}, \hspace{3pt}
$^e$ {\tt keisuke(at)th.phys.titech.ac.jp}, \hspace{3pt}
$^f$ {\tt walter.vinci(at)pi.infn.it}
}}

\begin{abstract}
We show that local/semilocal strings in Abelian/non-Abelian
gauge theories with critical couplings
always reconnect classically in collision, by using moduli space
approximation. The moduli matrix formalism explicitly identifies
a well-defined set of the vortex moduli parameters.
Our analysis of generic geodesic motion in terms of those
shows right-angle scattering in head-on collision of two vortices,
which is known to give the reconnection of the strings.
\end{abstract}

\maketitle


\noindent
{\bf Introduction.} ---
The issue of reconnection (intercommutation, recombination)
of colliding cosmic strings attracts much
interest recently
(see \cite{Hindmarsh:1994re,Jeannerot:2003qv,Urrestilla:2004eh}),
owing to the fact that the reconnection probability is related
to the number density of the cosmic strings, which is strongly
correlated with possible observation of them. However,
solitonic strings may appear in numerous varieties of field theories,
which certainly makes any prediction complicated. In this Letter, we
employ the moduli matrix formalism \cite{Eto:2006pg} to show that, in
a wide variety of field theories admitting supersymmetric
generalization,
inevitable reconnection of colliding solitonic strings
({\it i.e.} reconnection probability is unity) is universal.
The inevitable reconnection of local strings
in Abelian Higgs model \cite{Copeland:1986ng} (see
also \cite{Hanany:2005bc}) has been known for decades, and
for
non-Abelian local strings in $U(N_{\rm C})$ gauge theories
with $N_{\rm F}(=N_{\rm C})$ flavors,
this
universality was found in \cite{Hashimoto:2005hi} by a topological
argument. Here,
via a different logic and explicit computations,
we show the concrete dynamics of the inevitable reconnection
(note that \cite{Hashimoto:2005hi} does not describe dynamics).
Furthermore, our results extend the universality to
semilocal strings \cite{Vachaspati:1991dz} ($N_{\rm C}<N_{\rm F}$),
which is consistent with recent numerical simulations
\cite{Laguna:2006qr} (\cite{Achucarro:2006es}).
Stable semilocal strings are realistic and generic in many
supersymmetric
grand unified theories
\cite{Jeannerot:2003qv} and cosmologies \cite{Urrestilla:2004eh}.

The reconnection of the vortex strings can be understood
\cite{Copeland:1986ng} as
right-angle scattering of vortices in head-on collisions
\cite{vortex-collision} appearing in a spatial slice.
We use moduli space approximation where the motion of the strings
is slow enough, to find universal right-angle scattering of vortices
on two spatial dimensions.
The moduli matrix formalism \cite{Eto:2006pg}
gives a well-defined set of moduli
coordinates, and with that the analysis of the motion is quite simple
and robust.
Our results will be a basis for further analyses
on coupling to gravity and application to cosmology, and possible
comparison against cosmic super/D-strings
\cite{Jones:2002cv,Jackson:2004zg,Hanany:2005bc}.


\vspace{5pt}\noindent
{\bf Non-Abelian vortices.} ---
We deal with $U(N_{\rm C})$ gauge theory coupled to
$N_{\rm F}$ Higgs fields
$H$ ($N_{\rm C}\times N_{\rm F}$ matrix)
in the fundamental representation. Its Lagrangian is
\begin{eqnarray}
 {\rm Tr}
\left[
-\frac{1}{2g^2} F_{\mu\nu}F^{\mu\nu} + {\cal D}_\mu H
({\cal D}^\mu H)^\dagger-\frac{g^2}{4}
\left(c{\bf 1}_{N_{\rm C}}-HH^\dagger\right)^2
\right].
\nonumber
\end{eqnarray}
The Higgs self-coupling is put equal to the gauge coupling $g$
(critical coupling)
so that the theory admits supersymmetric extensions.
In the following we set $c>0$ to ensure stable vortex configurations.
The vortex equations
for strings extending along the $x^3$-axis are
\begin{eqnarray}
 {\cal D}_{\bar{z}} H=0, \quad
F_{12}+\frac{g^2}{2}\left(c{\bf 1}_{N_{\rm C}}-HH^\dagger\right)=0,
  \label{eq:vortex-eq}
\end{eqnarray}
where $z\equiv x^1 + ix^2$.
$k$ vortex solutions saturate the Bogomol'nyi energy
bound ${\cal E} \ge 2\pi c k$.
The moduli matrix formalism provides a method to identify moduli
(collective coordinates) of the solitons
and to describe the dynamics of the solitons by
collective motion.
Once the moduli matrix $H_0(z)$
which is an $N_{\rm C}$ by $N_{\rm F}$ holomorphic matrix with respect
to $z$
is given, one can solve the equations (\ref{eq:vortex-eq})
as \cite{Isozumi:2004vg,Eto:2005yh,Eto:2006pg}
\begin{eqnarray}
 H = S^{-1} H_0(z), \;
 A_1 + iA_2 = -2i S^{-1}\bar{\partial}_z S, \\
 \partial_z (\Omega^{-1}\bar{\partial}_z \Omega)
 = \frac{g^2}{4}\left(c{\bf 1}_{N_{\rm C}}
-\Omega^{-1}H_0H_0^\dagger\right),
 \label{mastereq}
\end{eqnarray}
where $S(z,\bar z)$ takes value in $GL(N_{\rm C},{\bf C})$ and
$\Omega \equiv S(z,\bar{z})S^\dagger (z, \bar{z})$ is
a gauge invariant quantity.
Equation (\ref{mastereq}), called the master equation,
is assumed to allow the unique and smooth solution for
any given $H_0$.
(This was rigorously proven for the cases of the Abelian gauge group and
of vortices on Riemann surfaces.
For general case of vortices on ${\bf C}$,
it is consistent with the index theorem \cite{Hanany:2003hp}).
Elements of $H_0$ are polynomial functions of $z$
and their coefficients are nothing but the moduli parameters.
The moduli space of the solitons is parameterized by these moduli.
The degree of det$(H_0H_0^\dagger)$ equals
the vortex number $k$. In this
Letter we use $k=2$ for describing collision of two vortex strings.
We need to fix the $V$-equivalence relation
$\{S(z,\bar z),H_0(z)\} \sim \{V(z)S(z,\bar z),V(z)H_0(z)\}$
with $V(z)\in GL(k,{\bf C})$
to get rid of unphysical redundancy.
After this fixing, the moduli matrix $H_0$
including $2kN_{\rm F}$ independent parameters
corresponds, by one-to-one, to a physical configuration.

The universal reconnection is shown based on
the fact that the moduli parameters
linear in $H_0$ (see (\ref{modulimatrix}) below) cover
the whole moduli space only once.
This ensures that our analysis is generic and
that the moduli metric is smooth and non-vanishing.
The K\"ahler potential
of the effective theory of the moduli parameters
was derived in \cite{Eto:2006uw}:
\begin{eqnarray}
 K = {\rm Tr}\int\!\! d^2z \left(
c\log\Omega+ c^{-1}\Omega^{-1} H_0 H_0^\dagger
+ {\cal O}(1/g^2)
\right).
\label{kahler}
\end{eqnarray}
This K\"ahler potential can be thought of as an
action functional for $\Omega$:
the equation of motion for $\Omega$, $\delta K/\delta \Omega=0$,
is identical with the master equation (\ref{mastereq}).
The smoothness of the solutions guarantees
the smoothness of the K\"ahler potential
and the absence of ultra-violet divergence.
Infra-red divergence of (\ref{kahler})
can exist
as non-normalizable modes,
which will be discussed later.
The one-to-one correspondence between $H_0$ and
the physical configurations implies
non-vanishing metric in terms of well-defined parameters.


\vspace{5pt}
\noindent
{\bf Reconnection of non-Abelian local strings.} ---
We deal with the local strings ($N_{\rm C}=N_{\rm F}$),
followed by the semilocal strings ($N_{\rm C}<N_{\rm F}$).
We will find that essential feature can be captured in the case
$N_{\rm C}=N_{\rm F}=2$.
Single vortex ($k=1$) moduli space is ${\bf C} \times {\bf C}P^1$
with ${\bf C}$ the position of
the vortex string in $z$-plane and ${\bf C}P^1$
the orientational
moduli concerning the internal color-flavor space
\cite{Hanany:2003hp,Auzzi:2003fs},
while the moduli space of separated two ($k=2$) vortices
is a symmetric product
$({\bf C} \times {\bf C}P^1)^2/{\mathfrak S}_2$.
The reconnection problem is related to
how they collide in the full $k=2$ moduli space,
parameterized by the moduli matrices \cite{Eto:2005yh}
\begin{equation}
H_0^{(0,2)}\! =
\left(
\begin{array}{cc}
1 & -az-b \\
0 & \!z^2\!-\!\alpha z\! -\! \beta
\end{array}
\right)
,
H_0^{(1,1)} \!=
\left(
\begin{array}{cc}
     \!z \!-\! \phi & \!\!-\eta \\
\!-\tilde{\eta} & \!\!\!z\!-\!\tilde{\phi}
\end{array}
\right).
\label{modulimatrix}
\end{equation}
The superscripts label patches covering the moduli
space, but one more patch (2,0) is needed to cover the whole manifold.
Since the (0,2) patch covers all the moduli space except lower-dimensional
submanifolds, this is sufficient for computing the reconnection
probability.
The moduli space of the two
coincident vortices in this theory has been studied in
\cite{Hashimoto:2005hi,Auzzi:2005gr,Eto:2006cx} and found
to be ${\bf C} \times W{\bf C}P^2_{(2,1,1)} \simeq
{\bf C}\times {\bf C}P^2/{\bf Z}_2$,
which any collision of strings goes through.
The locations $z_1$ and $z_2$ of the vortices
and the orientation vectors $\vec{\phi}_1$ and
$\vec{\phi}_2$ of the internal moduli for each vortex
are determined by
\begin{eqnarray}
 \det H_0 = (z-z_1)(z-z_2), \quad
H_0(z=z_i) \vec{\phi}_i =0.
\end{eqnarray}
We parameterize the vectors as
$\vec{\phi}_i = (b_i,1)^{\rm T}$
with $b_i = a z_i + b$, and the relations to
the original parameters are
\begin{eqnarray}
a = \frac{b_1\!-\!b_2}{z_1\!-\!z_2}, \;
b = \frac{b_2 z_1\! -\! b_1 z_2}{z_1\!-\!z_2},
\;
\alpha=z_1\!+\!z_2,\;
\beta = -z_1 z_2.
\label{relationzb}
\end{eqnarray}
Physical meaning of the parameters
$(z_i,b_i)$ is clear, but
they can cover only the subspace $z_1\not =z_2$
because the relations (\ref{relationzb})
are not defined at $z_1 =z_2$.

Let us consider slow motion of the moduli parameters,
{\it \`a la} Manton \cite{Manton:1981mp}, to show the
universal right-angle scattering in the vortex collision.
We have to use the parameters
$(a,b,\alpha,\beta)$, not $(z_i,b_i)$, because,
as we have shown, the moduli space metric with respect to the
former parameters (which appear linearly in the moduli matrix $H_0$)
is smooth and non-vanishing.
With these ``well-defined'' parameters of
the moduli space, at least for a certain period of time
around the collision moment,
one can approximate the moduli motion as
linear functions of $t$
(since the coordinates are subject to free motion):
\begin{eqnarray}
&&a = a_0 + \epsilon_1 t + {\cal O}(t^2), \quad
b = b_0 + \epsilon_2 t + {\cal O}(t^2),
\label{modulimotion1} \\
&&\alpha= 0 + {\cal O}(t^2),\quad
\beta = \epsilon_3 t + {\cal O}(t^2),
\label{modulimotion2}
\end{eqnarray}
where $\epsilon_i$, $a_0$ and $b_0$ are constant.
Here $\alpha$ is the center of mass of the vortices (see the later
discussion for identifying the decoupled center-of-mass parameter),
and thus set to be zero around $t=0$. We have used a time translation
so that a constant term in $\beta(t)$ vanishes.
This is equivalent to choose the collision moment as $t=0$.

Physical interpretation of the motion (\ref{modulimotion1}) and
(\ref{modulimotion2})
can be extracted by looking at the solution in terms of $z_i$
and $b_i$. From (\ref{relationzb}), we obtain
\begin{eqnarray}
& z_1 = -z_2 = \sqrt{\epsilon_3 t} + {\cal O}(t^{3/2}),&
\label{z1z2re}\\
& b_i = b_0 + (-1)^{i-1} a_0\sqrt{\epsilon_3 t} + {\cal O}(t).
&
\label{bicol}
\end{eqnarray}
The first equation shows that the vortices are scattered by the right
angle; since the time dependence is $\sqrt{t}$, when time
varies from negative to positive, the vortex moves from
the imaginary axis to the real axis.
As stressed before, this right-angle scattering means that the vortex
strings are reconnected. So, generic collision results always in
reconnection.

When $a_0=0$ in (\ref{bicol}), the orientational moduli for each
vortex coincide, which corresponds to a reduction
to the case of the Abelian-Higgs model. Here we have shown that even
when $a_0\neq 0$ and the non-Abelian
strings have different orientational moduli at the initial time, as they
approach each
other in the real space, the internal moduli approach each other; in
particular,  $b_i$ experiences the right-angle scattering, too.
This is the only consistent solution to the moduli equations of
motion, with generic initial conditions.
Note that this understanding
comes from the re-description in terms of $b_i$ and $z_i$, while
the true and correct motion in the moduli space is determined by
the moduli parameters $(a,b,\alpha,\beta)$, which have linear
dependence in $t$.

Although we have shown (by using the (0,2) patch) that the reconnection
probability is unity, it is instructive to look at the other patches to
see what happens in the submanifold(s) of the moduli space which cannot
be described by the (0,2) patch. In fact the submanifold includes
the ${\bf Z}_2$ singularity of the
${\bf C}P^2/{\bf Z}_2$.
This corresponds
to the situation where the vortices sit in two decoupled $U(1)$
sub-sectors of the $U(2)$ in the original field theory
and where strings should pass through each other in
collision in that special case.
In the (1,1) patch,
the condition for coincident vortices, namely $\det H_0=z^2$,
reads
\begin{eqnarray}
\tilde\phi = -\phi, \quad
\phi\tilde\phi - \eta\tilde\eta = 0,
\label{phicon}
\end{eqnarray}
which can be parameterized by $X$ and $Y$ through
$ XY = -\phi = \tilde\phi, X^2 \equiv \eta, Y^2 \equiv -\tilde\eta$.
The ${\bf Z}_2$ symmetry $(X,Y)\sim(-X,-Y)$ is manifest
\cite{Eto:2006cx}. Note that the orbifold singularity
$X\!=\!Y\!=\!0$
($\eta\!=\!\phi\!=\!\tilde\eta\!=\!\tilde\phi\!=\!0$) is present
only in the submanifold $z_1=z_2$, while the full moduli space is
smooth.
One can confirm this by computing the K\"ahler potential
explicitly around the origin of the (1,1) patch,
$K = 2\pi c (|\phi|^2\! +\! |\tilde\phi|^2
\!+\! |\eta|^2\! +\! |\tilde\eta|^2)\!+\! \mbox{higher}$,
which shows that there the metric
is smooth and non-vanishing.
Going to the $(X,Y)$ coordinates on the submanifold,
we obtain a metric of a ${\bf Z}_2$ orbifold,
$K \propto (|X|^2 + |Y|^2)^2$.

Let us study geodesic motion on the moduli space
to see the reconnection.
After imposing the center-of-mass condition $z_1=-z_2$,
we obtain the motion of the moduli parameters
\begin{eqnarray}
&& \phi = -\tilde\phi = - XY+s_1 t
+ {\cal O}(t^2),
\label{solxy1}\\
&& \eta =X^2+s_2 t + {\cal O}(t^2),\quad
 \tilde\eta =- Y^2+ s_3 t
+ {\cal O}(t^2),\qquad
\label{solxy2}
\end{eqnarray}
where $X, Y$ and $s_{1,2,3}$ are constant.
We have chosen the collision moment to be $t=0$,
so that the constant terms in the above satisfy the constraint
(\ref{phicon}).
The orientational moduli $b_i$ are
obtained as $b_i=\eta/(z_i-\phi)$.

From this generic solution
of the equations of motion, we compute
(for $|X|^2+|Y|^2\neq 0$)
\begin{eqnarray}
&& z_1 =-z_2= \sqrt{\phi^2 + \eta\tilde\eta}=
\sqrt{st} + {\cal O}(t^{3/2}),
\\
&& b_i=X Y^{-1} + (-1)^i Y^{-2} \sqrt{st} +{\cal O}(t),
\label{bicol2}
\end{eqnarray}
where $s \equiv -2s_1XY+s_3X^2-s_2Y^2$.
Therefore, we confirm the generic reconnection for $s\neq 0$.
The condition $s=0$ is equivalent to
$\epsilon_3=0$ in the analysis of the (0,2) patch, because
among the patches we have a relation
$\beta = \eta\tilde\eta - \phi \tilde\phi = st$.
$s=\epsilon_3=0$ can be achieved only by finely tuned initial
conditions, so we are not interested in it.

When $X=Y=0$ (this point is not covered by the (0,2) patch, so
the identification $s = \epsilon_3$ fails), we obtain
\begin{eqnarray}
&& z_1 =-z_2=
\sqrt{s_1^2 + s_2 s_3} \;\; t + {\cal O}(t^{3/2})
\label{solxy=0},\\
&& b_i=s_1s_3^{-1} + (-1)^{i-1}s_3^{-1}\sqrt{s_1^2+s_2s_3}
+{\cal O}(t^{1/2}),\quad
\end{eqnarray}
which shows  no reconnection.
Note that this finely tuned collision allows constant
non-parallel orientations $b_1\neq b_2$ at the collision,
in contrast to the general case (\ref{bicol}) (\ref{bicol2})
where $b_1=b_2$ at $t=0$.
One observes that the reconnection is intimately related to the
parallelism of the orientation vectors $b_i$, as is along the intuition.
But the significant is that parallel $b_i$ at the collision moment
follows from generic initial conditions, which is  clarified here in
the explicit computations in the moduli matrix formalism.

For $N_{\rm C}\!=\!N_{\rm F}>2$ (the orientational moduli space
is ${\bf C}P^{N_{\rm C}-1}$),
the same argument finds that the probability is unity.
The moduli matrix of $(0,\cdots,0,2)$ patch is
\begin{equation}
 H_0^{(0,\cdots,0,2)} =
\left(\begin{array}{cc}
{\bf 1}_{N_{\rm C}-1} & \vec{a}z-\vec{b} \\
\vec{0}\, {}^{\rm T} &  z^2-\alpha z -\beta
\end{array}
\right)
.
\end{equation}
The center-of-mass parameter is identified with $\alpha$ and we put it
zero. Then, we have $\beta = z_1^2$, and the solution of
the equation of motion for $\beta$ is the same as (\ref{modulimotion2}),
after the time translation.
Finally we have (\ref{z1z2re}), therefore we conclude that reconnection
occurs, irrespective of the other moduli parameters $\vec{a}$ and
$\vec{b}$. Because the (0,0,$\cdots$,2) patch covers generic points of
the moduli space, the reconnection probability is unity.
The results are completely consistent with
\cite{Hashimoto:2005hi} which used a different logic though.


\vspace{5pt} \noindent {\bf Reconnection of semilocal strings.} ---
We shall show that the reconnection probability is unity also for
the semilocal strings, $N_{\rm C}<N_{\rm F}$. We follow the same
logic and find that it applies to rather generic theories, showing
universality of reconnection.
It is enough to show this for the simplest example with
$N_{\rm C}=2$ and $N_{\rm F}=3$.
The moduli matrix is \cite{Eto:2006pg}
\begin{eqnarray}
 H_0 =
\left(
\begin{array}{ccc}
1 & -az-b & -ad \\
0 & z^2-\alpha z -\beta & dz + e
\end{array}
\right).
\label{h0semi}
\end{eqnarray}
In the following, we shall show that (i) even in this semilocal case
the center-of-mass coordinate is $\alpha$ and thus put to be zero, and
(ii) the parameter $d$ (which is associated with the size of the vortex)
and the combination $bd + ae + ad\alpha$
are non-normalizable and put to be constant.
Using these facts, the logic leading to the reconnection is the same
for the remaining normalizable parameters:
$z_1= \sqrt{\beta} = \sqrt{\epsilon_3 t}$.
We find the universality in reconnection.
Note that the additional moduli parameters appearing from the extra
flavors, $d$ and $e$, does not play any role in showing the
reconnection. This is clearly the same for more general non-Abelian
semilocal strings.
With the help of the moduli matrix, one can also show that the
reconnected
semilocal strings have the same width, which is expected from a
geometrical viewpoint.

Let us identify the non-normalizable modes by studying possible
infra-red divergence in the K\"ahler potential (\ref{kahler}).
The asymptotic boundary condition for the master equation
(\ref{mastereq}) is $ \Omega \to (1/c) H_0 H_0^\dagger$, and using the
expression of $H_0$ (\ref{h0semi}), we find only the first term in
(\ref{kahler}) is relevant.
After a K\"ahler transformation,
$K$ is evaluated for large $|z|$,
\begin{eqnarray}
&&K\sim c \int\!\! d^2z \log
\left[1+ \frac{|d|^2+|bd+ae+ad\alpha|^2}{|z|^2}\right]
\nonumber \\
&&\sim
c \int\!\! d^2z \frac{|d|^2+|bd+ae+ad\alpha|^2}{|z|^2}
\nonumber \\
&&\sim
2\pi c (|d|^2+|bd+ae+ad\alpha|^2) \log L
\end{eqnarray}
where
in the last expression
we introduced a cut-off radius
$L(\to\infty)$.
This divergence shows that the parameter $d$
and the combination $bd + ae+ad\alpha$ are
non-normalizable.
We have to fix these modes to be constant, so that the effective
Lagrangian is finite. In other words, motion of these parameters
is frozen because the kinetic term of those diverges and any motion costs
infinite energy. Other parameters are normalizable, oppositely to the single vortex case \cite{Shifman:2006kd}.

Next, we provide a method to determine the center-of-mass parameter,
which is decoupled from the others.
We write the moduli matrix in the following form,
\begin{eqnarray}
 H_0 =
\left(
\begin{array}{ccc}
1 & -a(z-z_3) & -ad \\
0 & (z-z_1)(z-z_2) & d(z-z_4)
\end{array}
\right)
\label{hzeroz}
\end{eqnarray}
in which the parameters are not the ``well-defined'' parameters.
In this form, there is a translation symmetry
$ z \to z+\delta, z_i \to z_i + \delta$.
Let us assume that $z_0$, which is a linear combination of $z_1$, $z_2$,
$z_3$ and $z_4$, is the center-of-mass parameter.
The other two parameters independent of $z_0$ should be selected properly
from the three  $z_i' \equiv z_i-z_0 \; (i=1,2,3,4)$.
We compute the metric from the K\"ahler potential, for this set of
independent coordinates.
The complete decoupling of $z_0$ from the remaining parameters is
ensured if the metric component
$g_{i\bar{0}}\equiv \delta^2 K/\delta z'_i \delta \bar{z}_0$
vanishes. We can compute it as
\begin{eqnarray}
 g_{i\bar{0}}
&=&
-\frac{\delta}{\delta z'_i}
\int\!\! d^2z \frac{\delta }{\delta \bar{z}}
\tilde{K}(z,z_0,z_j')
=
-\frac{\delta}{\delta z'_i}
\oint\! dz
\tilde{K}, \;\;\;
\end{eqnarray}
where $\tilde{K}$ is the integrand of the K\"ahler potential,
and we used the fact that $z_0$ dependence in
$\tilde{K}$ is always through the combination $z-z_0$.
The explicit expression (\ref{hzeroz}) gives, after an appropriate
K\"ahler transformation, for large $|z|$,
\begin{eqnarray}
-\frac{\delta}{\delta z'_i}\!
\oint\!\! dz
c\log \left(
\!1\! -\! \frac{z_1\!\!+\!\!z_2}{z}\! -\!\frac{\bar{z}_1\!\!+\!\!
\bar{z}_2}{\bar{z}}\! +\! \cdots
\!\right)
=4\pi c \frac{\delta}{\delta z'_i}\frac{z_1\!+\!z_2}2.
\nonumber
\end{eqnarray}
Vanishing of this means that $z_i'$ is orthogonal to the
combination $z_1+z_2$, which shows that the center-of-mass parameter is
$z_0 = (z_1 + z_2)/2 = \alpha/2$.
This result is non-trivial, because there
are other dimensionful parameters $z_3$ and $z_4$ which might have been
involved with the definition of the center-of-mass.


\vspace{5pt}
\noindent
{\bf Conclusions.} ---
The moduli matrix formalism has shown that local/semilocal strings in
Abelian/non-Abelian gauge theories with critical couplings always
reconnect classically in collision.

While we studied the critical coupling in this Letter,
non-critical region
(which can be smoothly deformed from the critical coupling)
has the same universality,
since in the moduli space it is described by introduction
of potential terms along relative position moduli induced by
attractive/repulsive force between type I/II strings.
Even for the repulsive case two strings must collide,
because 
parts of two strings far from the collision point
do not feel a force and
the potential induced around the collision point
is negligible compared with the total string energy.
Adding small mass terms breaking flavor symmetry can be
treated similarly (see for example \cite{Hashimoto:2005hi}).

The universal reconnection found in this Letter uses
the moduli space approximation, and is valid
below the energy scale of the first massive excitation in the soliton
background. In the case that collision speed exceeds this limitation,
one needs incorporation of the massive modes.
As in the case of Abelian-Higgs model, numerical
simulations \cite{Laguna:2006qr,Achucarro:2006es} showed robustness of the reconnection
even for high energy collisions.
We hope that, in the future
observation, this universality may help for distinguishing solitonic
strings from cosmic superstrings/D-branes which have lower
reconnection probabilities \cite{Jackson:2004zg,Hanany:2005bc}.
The moduli matrix formalism has opened up new paths to analyze BPS
solitons. It would be intriguing to apply it further to more
involved/realistic situations, such as
cosmic string webs and thermal phase transitions.

\vspace{5pt}
Acknowledgments:
K.H.~would like to thank K.~Konishi for his hospitality.
G.M. wants to thank J.~Baptista and N.~Manton for useful conversations.
M.E., M.N. and K.O would like to thank the theoretical HEP group of KIAS
and to Erick Weinberg for useful comments.
G.M.~and W.V.~want to thank the Theoretical HEP Group of the Tokyo
Institute of Technology for their warm hospitality.
The work of M.E.~and K.O~is supported by Japan Society for the Promotion
of Science (JSPS) under the Post-doctoral Research Program.
K.H.~is partly supported by JSPS and
the Japan Ministry of Education, Culture, Sports, Science and
Technology. G.M.~acknowledges the Foreign Graduate student Invitation
Program (FGIP) of the Tokyo Institute of Technology.

\vspace{-0.5cm}

\end{document}